# Time Dependent Density Functional Theory calculations of the optical properties of charge-transfer complexes

Satwik Ramanjanappa[a] and Edward R Van Keuren[b]

[a]Institut für Experimentalphysik und Zentrum für Quantenphysik, Universität Innsbruck, Technikerstraße 25, Innsbruck, 6020, Austria; [b] *Department of Physics, Institute for Soft Matter Synthesis and Metrology, and Center for Translational Imaging, Georgetown University, Washington, DC, USA

**ABSTRACT**

Charge transfer complexes are materials with a wide range of interesting optical and electronic properties. They have seen a great deal of research over the past decade, both in device development as well as research to elucidate the underlying mechanisms of charge transfer and transport. Here we present time-dependent density-functional theory (TD-DFT) calculations of the energy levels of the donor-acceptor complexes perylene-TCNQ (7,7',8,8'-tetracyanoquinodimethane) and hexamethylbenzene-chloranil. The calculations show a good match to the experimentally measured optical absorption spectra, and indicate that low energy CT absorption bands appear at the dimer/tetramer stage of cocrystal formation. We also report the degree of charge transfer for these complexes in various stoichiometries and compare them.

## 1. Introduction

Charge transfer (CT) cocrystals are an interesting class of materials with potential applications in optics and electronics [1]. These materials consist of molecular crystals containing both electron donating and accepting molecules, in which the partial transfer of charge from donor to acceptor generally leads to additional energy states. This can result in materials that are conducting [2], semiconducting [3], superconducting [4], ferroelectric [5], etc. The wide variety of possible donors and acceptors that can form CT cocrystals, and the ability to fabricate aligned CT fibers using scalable manufacturing methods [6], makes them attractive for numerous applications such as field effect transistors [7]. Various methods exist for inducing crystallization of CT cocrystals [8]. We have previously reported on the formation of several types of cocrystals using a method known variously as micronization [9], the reprecipitation method [10], or flash nanoprecipitation [11], in which a solution is rapidly mixed with a miscible non-solvent. The rapid drop in solubility causes self-assembly of the solute molecules, and at low concentrations these form metastable dispersions of nanocrystals. We have demonstrated the use of this method to synthesize cocrystals of perylene-TCNQ (TCNQ = 7,7',8,8',tetracyanoquinodimethane) [12] [13], hexamethylbenzene-chloranil [14] and phenothiazine-TCNQ [6].

CONTACT Edward R Van Keuren. Email: Edward.Van.Keuren@georgetown.edu

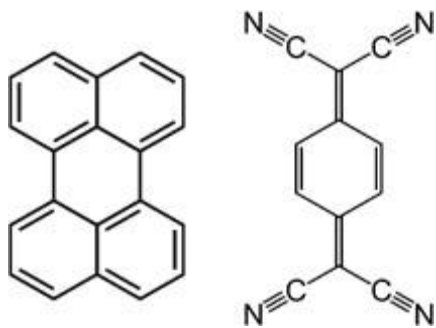

Figure 1.: Molecular structure of perylene (left) and TCNQ (right).

The optical and electronic properties of perylene-TCNQ (P-T) have been studied by a number of groups [15] [16] As shown in figure 1, perylene is a five-ringed aromatic molecule that is the base for many dyes, and TCNQ is a widely used molecule known for its strong electron accepting nature. The P-T cocrystal has been shown to have a number of possible stoichiometries: 1:1, 2:1 and 3:1, with at least two polymorphs reported for the 1:1 cocrystal [15][16]. The new charge-transfer states in cocrystals result in significant changes in the optical absorption spectra, described by Michaud several decades ago [17]. Their appearance has been used as a way to monitor the kinetics of the formation of nano-cocrystal formation as a function of the preparation conditions [13]. Hexamethylbenzene(HMB)-chloranil(CA) is another well-known charge transfer crystal, having been studied since the 1950's [22] Similar to many other CT cocrystals, HMB and CA assemble in an alternating stack, with the appearance of a charge-transfer absorption peak around 540 nm. This new peak was used to assess the effectiveness of an impinging jet mixer for creating cocrystals [14]. In order to understand the nature of the charge-transfer and its effect on the electronic and optical properties, many groups have employed density functional theory (DFT) calculations[18]. For example, the electronic band structure of several stoichiometries of perylene-TCNQ cocrystals were determined by Vermeulen et al[16]. While DFT in general yields accurate values for ground state energy levels, time-dependent DFT (TDDFT) enables the determination of excitation energies and thus the optical absorption spectra of materials [19][20].

Here we present TDDFT calculations of the donor-acceptor complexes of P-T and HMB-CA and compare them to the measured absorption spectra of the cocrystals. The calculations also enable a determination of the degree of charge transfer from the donor to acceptor in the different complexes, which we compare to the data found from Raman spectra using the method for TCNQ reported by Matsuzaki [21]. We also calculate the absorption spectrum of the 2:1 P-T cocrystal, which has not yet been reported experimentally.

## 2. Computational methods

Time-Dependent Density functional Theory (TD-DFT) is a powerful theoretical framework that extends the principles of Density Functional Theory (DFT) to investigate various dynamical and electronic properties of materials over time. Detailed description of DFT and TD-DFT can be found in various standard textbooks [35]. A recent breakthrough involves the introduction of simplified versions of Time-Dependent

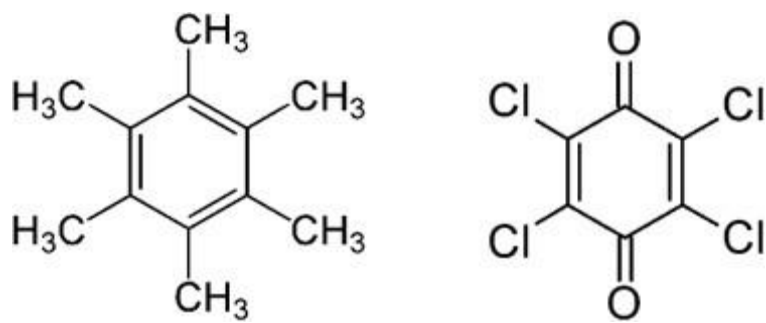

Figure 2.: Molecular structure of hexamethylbenzene (left) and chloranil (right).

Density Functional Theory (sTD-DFT) and the simplified Tamm-Dancoff Approximation (sTDA), which were pioneered by Grimme et al [34]). These advancements align with the contemporary direction of developing quantum-chemical techniques that are accessible for handling sizable molecules using moderate computational resources, including standard desktop computers. This innovation has made it possible to perform sophisticated calculations on large molecular systems without the need for extensive computational infrastructure. This represents a significant step forward in making quantum-chemical research more feasible and widely applicable. We implemented sTD-DFT using the ORCA 5.0.4 [23][24] software package to investigate the charge transfer properties of several organic molecules in different stoichiometries and configurations. We use the generalized gradient approximation functional (GGA) BP86 [26][27] to perform the TD-DFT calculations. A very simple BP86 functional along with double basis sets def2-SVP was used because it is known to provide accurate geometries for large systems with a greater number of atoms or cycles [38][39]. To improve the accuracy and computation time, we choose an optimum compromise between the maximum dimension of the Davidson matrix and the expected number of excitations to be resolved. We studied these complexes in the solvents acetone and ethanol. We first obtained geometrical optimized structures for each of the molecules and complexes via Avogadro software [28], followed by generation of the required ORCA input files to perform the TD-DFT calculations. To invoke the influence of solvents on these complexes we used the CPCM(solvent) command within the TD-DFT functional [25][40], which introduces a continuum polar solvent into the simulation. We obtained the UV-Vis-NIR spectra of the complexes by calling the output file obtained from TD-DFT calculations and calculating the various excitations using ORCA's executable *orcamapspc*. In the following section we present the UV-Vis-NIR spectra of the complexes in various stoichiometries that we obtained from these calculations and compare them to the experimental results.

## 3. Results

The calculated optical absorption spectrum of the individual perylene and TCNQ molecules in acetone in the UV-Vis-NIR range are shown in Figure 3. The spectrum of neutral TCNQ has a sharp prominent peak around 400 nm and agrees well with the experimental results reported in the literature [29] [30]. UV-Vis-NIR spectrum of perylene shows large peaks close to 300 nm and 460 nm along with smaller peaks around 250 nm and 330 nm [31]. The positions of these peaks were also seen in an experimental study [32] in cyclohexane solvent. The effect of the solvent is known to be very negligible and at most shifts the peaks positions by tens of nm [33].

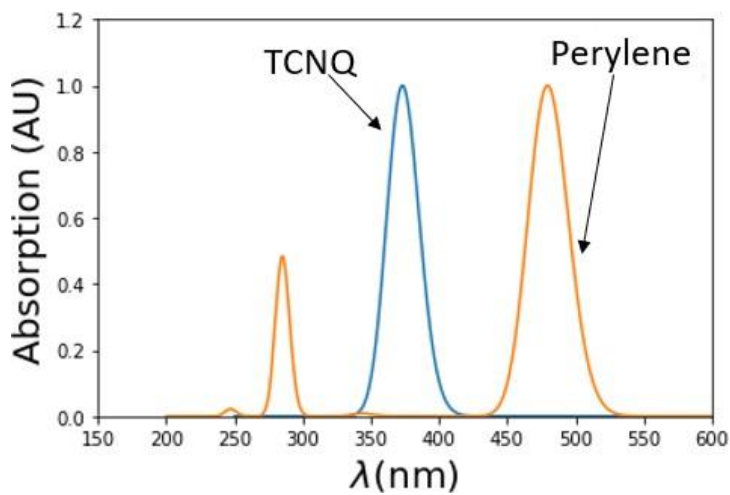

Figure 3.: Calculated UV-Vis optical absorption spectra of perylene and TCNQ.

### *3.1. P1T1 single and stacked dimers*

The optimized geometry of the 1:1 P-T complex (P1T1) was obtained using Avogadro and TD-DFT calculations were used to obtain the absorption spectra. Figure 4 displays the absorption spectrum of P1T1, obtained both experimentally in nanococrystals [12] [29] and on P1T1 dimers using TD-DFT. The peaks due to excitations localized in the individual perylene and TCNQ molecules are observed around 450 nm but cannot be resolved as they fall very close to each other. The main feature in the experimental spectrum is the charge transfer peak at approximately 1040 nm. This peak is accurately reproduced in the calculated spectrum. Li, et al. also reported a small peak around 910 nm, which is characteristic of the 1:1 P-T complex and was not observed in other stoichiometries such as P3T1 [12]. Our TD-DFT calculations also reproduced this peak, as seen in the figure. However, we also calculated a very strong peak near 750 nm which is possibly due to the anionic TCNQ, which was not clearly observed in the experimental data at the same position [12]. A strong peak is not observed in the experimental spectrum at this wavelength, although a small bump can be seen in the spectrum near 840 nm.

The delocalization of electrons in conjugated materials can lead to significant changes in the absorption spectrum [36], so in order to gain a better comparison with the experimental spectra measured from cocrystals, we looked at the effect of stacking multiple dimers on the calculated spectrum. Figure 5 shows the calculated spectra of the P-T complex and a stack of two dimers. The 1040 nm peak broadens significantly and become the predominant feature, while the shorter wavelength peaks attributed to the molecular energy levels in perylene and TCNQ alone decreased in intensity. This matches the experimental data, where the 1040 nm peak increased and the pure component peaks around 450 nm decreased over time as the nanocrystals grew [12]. This enhancement in the oscillator strength of the CT state is likely due to the increase in intermolecular charge delocalization.

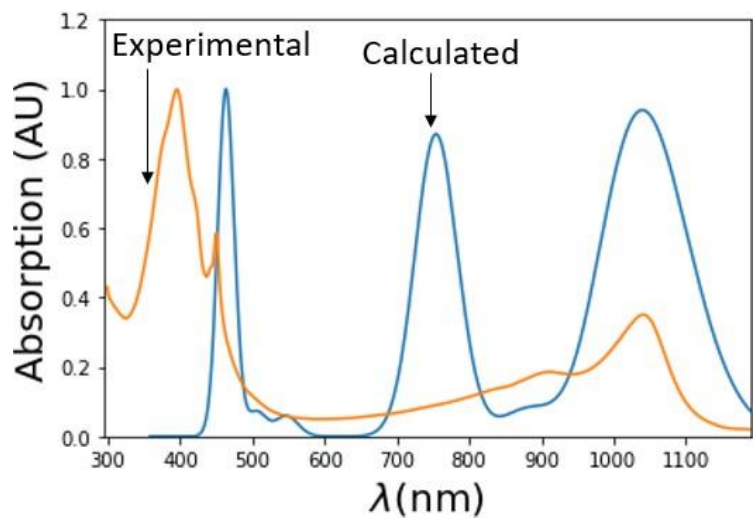

Figure 4.: The calculated and experimental absorption spectra of a single P1T1 dimer.

### 3.2. P3T1

In a similar manner, we investigated the 3:1 P-T complex (P3T1). As with the P1T1 complex, we observed the low-wavelength peaks corresponding to pure TCNQ and perylene, as shown in Figure 6. Additionally, we found a charge transfer (CT) peak around 980 nm, which aligns with the experimental findings, where a similar peak was observed around 960 nm [12][17]. Furthermore, we observed a smaller, broader peak near 800 nm. However, it remains unclear whether this peak corresponds to pure TCNQ or anionic TCNQ. It's possible that this peak contains information from both configurations, as it appears to span a wide range of wavelengths, making it challenging to resolve smaller peaks. When compared to experimental data we see a remarkable agreement on the position of CT peak, while the peaks corresponding to the TCNQ and perylene molecules were observed to have shifted negligibly by around 20 nm, which could be due to orbital overlaps between the molecules.

### 3.3. HMB-Chloranil

Experimentally determined spectra of pure HMB and chloranil in ethanol showed lowest energy absorption peaks of the individual components at 285 nm and 273 nm respectively [14]. The same was reproduced in our calculations for individual molecules differing by only a few nm as seen in Figure 7. These slight shifts could be due to solvation effects. We also calculated the absorption spectrum of the HMB-CA dimer, and compare it to the experimental results in Figure 8 [14][22]. The CT peak measured at around 550 nm was also seen in the simulated data from the TD-DFT runs. The peaks at shorter wavelengths correspond to peaks of individual species which could be due to extra polar and finite range effects in experimental samples leading to higher energy peaks. The CT peaks are an indication of charge transfer between the donor-acceptor pairs, the position of the CT peak is not seen to be affected much compared to the experimental data.

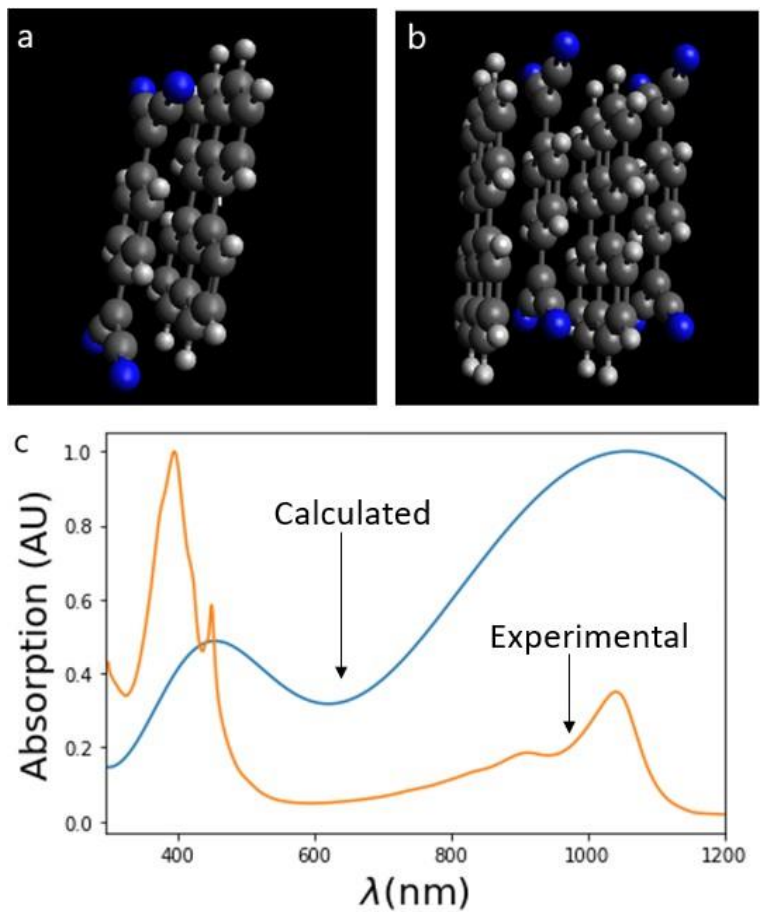

Figure 5.: (a) Molecular structure of the P1T1 dimer, (b) structure of two stacked P1T1 dimers, (c) calculated spectrum of the stacked dimers and experimental spectrum measured from P1T1 cocrystals.

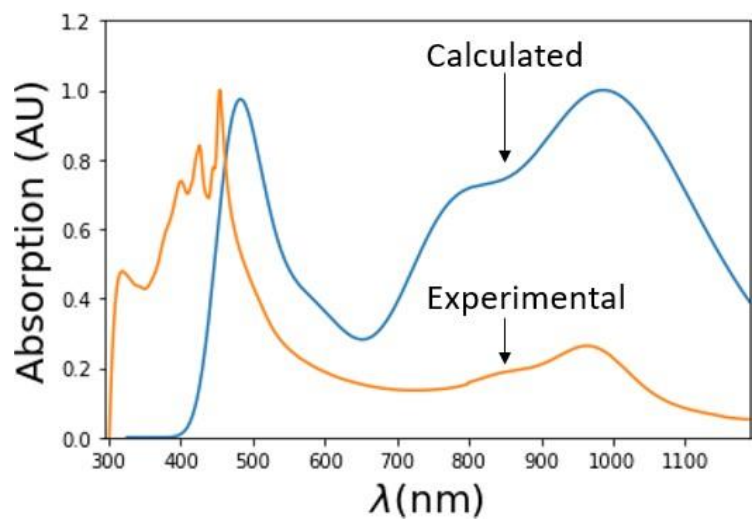

Figure 6.: Calculated and measured absorption spectra of P3T1.

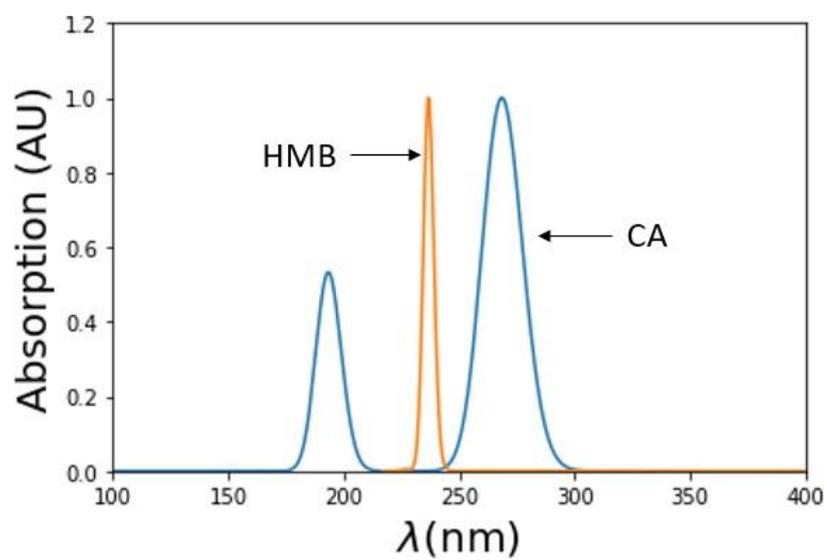

Figure 7.: Calculated UV-Vis-NIR absorption spectra of HMB and CA.

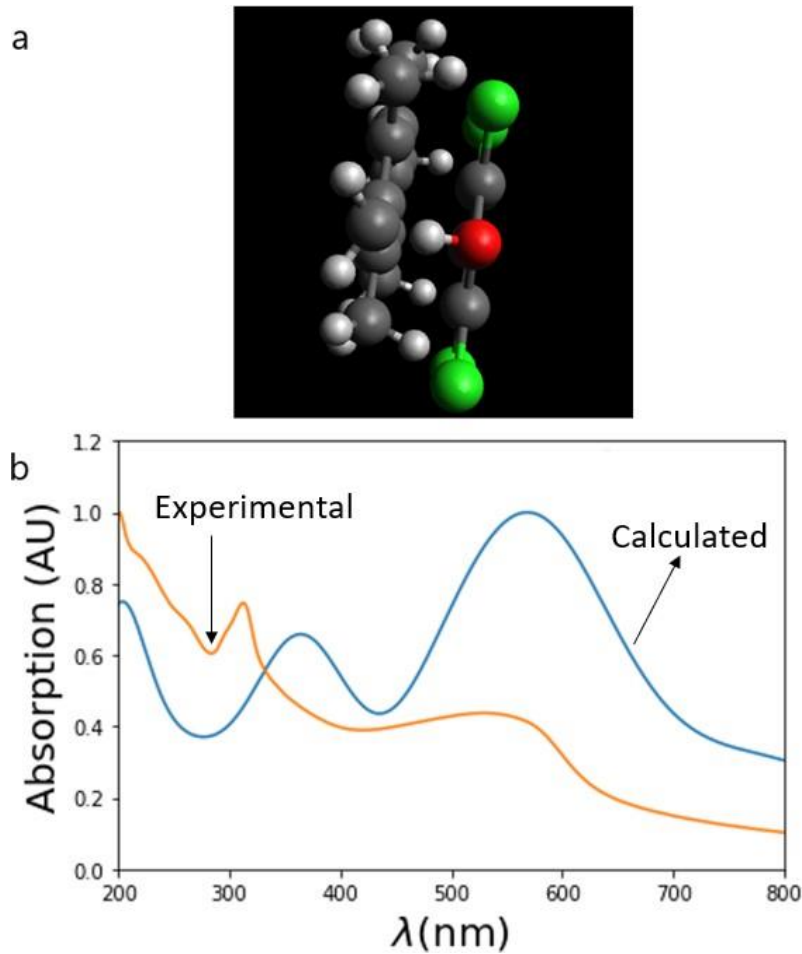

Figure 8.: (a) Molecular structure of the hexamethylbenzene:chloranil complex, (b) Calculated and experimental spectra.

### 3.4. Degree of charge transfer

A simple method suggested by Gao et al [37] was used to estimate the degree of charge transfer in the perylene-TCNQ complexes. This method leverages the observation that an increase in the electron density in the LUMO of TCNQ leads to a lengthening of the double bond (a decrease in bond order) in the TCNQ molecule (a in Figure 9) and a shortening of the single bonds adjacent to the double bond (b and c in Figure 9). The degree of charge transfer (DCT) was estimated using various quantities labeled $\alpha_{CT}$, $\alpha_0$, and $\alpha_{-1}$, which are functions of the bond orders $\alpha = a/(b + c)$, where the subscripts CT, 0, and -1 refer to the charge transfer crystal, neutral TCNQ crystal, and anionic TCNQ crystal, respectively. The degree of charge transfer (DCT) was then computed using DCT = $(\alpha_{CT} - \alpha_0)/ (\alpha_{-1} - \alpha_0)$. Values of DCT obtained are given in the Table 1.

| Crystal | $\alpha_0$ | $\alpha_{-1}$ | $\alpha_{CT}$ | DCT |
|---|---|---|---|---|
| P1T1 | 0.6637 | 0.5819 | 0.6649 | 0.0154 |
| P2T1 | 0.6637 | 0.5819 | 0.6683 | 0.056 |
| P3T1 | 0.6637 | 0.5819 | 0.6776 | 0.1702 |
| CA:HMB 1:1 | 0.3727 | 0.3692 | 0.3728 | 0.023 |

Table 1.: List of bond orders and DCT values for various complexes and their stoichiometries.

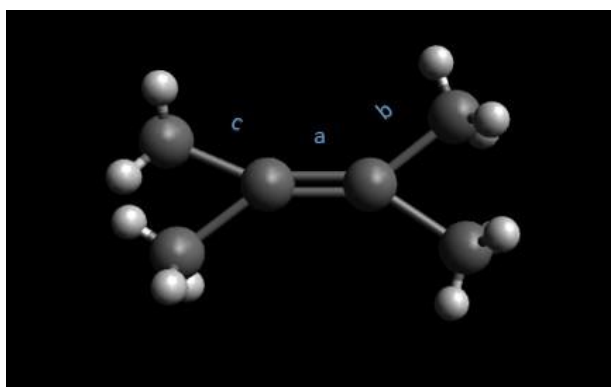

Figure 9.: A section of TCNQ labelling a, b and c.

The calculated value of DCT in the P-T complexes is given in Table 1. The bond orders are computed as part of the output of TD-DFT calculations for the excited states. Although this method may not provide accurate values for the degree of charge transfer, it could be a useful guide for comparing the relative values among different stoichiometries of the PT complex. For P1T1, we found the degree of charge transfer to be 0.0154, as calculated from the values of bond order as described above. This value seemingly agrees well with the values reported in Gao et al [37]. We further computed the DCT in the P2T1 stoichiometry using the same method and obtained 0.056. As expected, the DCT in P2T1 is larger in comparison to P1T1 and even more enhanced in P3T1 compared to both the other stoichiometries, with a DCT of 0.1702.

However, the values of the DCT are not in precise agreement with the values reported in Gao et al [37], but the trend of DCT between different stoichiometries is well-preserved, and the values lie within the reported experimental error. We further calculated the DCT of HMB-CA crystals using the same technique using the fact that Chloranil acts like an active acceptor (computing its anionic and neutral crystal properties) and HMB as a donor. With CA as the acceptor, we employ the same expressions as those utilized for the PT complex. We found the DCT in the HMB-CA complex in the 1:1 stoichiometry to be 0.023.

### 3.5. Conclusion

We have shown that TD-DFT can provide UV-Vis-NIR spectra of several charge transfer complexes in three different stoichiometries of perylene-TCNQ and hexamethylbenzene-chloranil. We found a close agreement between experimental and theoretical data for P1T1 and P3T1, with charge transfer peaks reported at 1040 nm and 980 nm, respectively. Furthermore, we compared the experimental and calculated data corresponding to the HMB-Chloranil complex, noting a good agreement between the charge transfer peaks. In addition, we observed changes in the calculated spectra going from the dimer complex to larger stacks of dimers, which was consistent with experimental findings and highlights the effect of delocalization of the excitations on the absorption spectra. Finally, the degree of charge transfer calculated from the TD-DFT results showed trends agreeing with the experimental results as the stoichiometry of the cocrystals was changed.